# Opening Band Gap without Breaking Lattice Symmetry: A New Route toward Robust Graphene-Based Nanoelectronics

Liangzhi Kou[†,*,#], Feiming Hu[†,*], Binghai Yan[‡,§,*], Thomas Frauenheim[†] and Changfeng Chen[$]

[†]Bremen Center for Computational Materials Science, University of Bremen, Am Falturm 1, 28359, Bremen, Germany

[‡] Max Planck Institute for Chemical Physics of Solids, NoethnitzerStr. 40, 01187 Dresden, Germany

[§]Max Planck Institute for Physics of Complex Systems, NoethnitzerStr. 38, 01187 Dresden, Germany

[$]Department of Physics and Astronomy and High Pressure Science and Engineering Center, University of Nevada, Las Vegas, Nevada 89154, United States

*These authors contributed equally   #Corresponding author: kouliangzhi@gmail.com

Developing graphene-based nanoelectronics hinges on opening a band gap in the electronic structure of graphene, which is commonly achieved by breaking the inversion symmetry of the graphene lattice via an electric field (gate bias) or asymmetric doping of graphene layers. Here we introduce a new design strategy that places a bilayer graphene sheet sandwiched between two cladding layers of materials that possess strong spin-orbit coupling (e.g., $Bi_2Te_3$). Our *ab initio* and tight-binding calculations show that proximity enhanced spin-orbit coupling effect opens a large (44 meV) band gap in bilayer graphene without breaking its lattice symmetry, and the band gap can be effectively tuned by interlayer stacking pattern and significantly enhanced by interlayer compression. The feasibility of this quantum-well structure is demonstrated by recent experimental realization of high-quality heterojunctions between graphene and $Bi_2Te_3$, and this design also conforms to existing fabrication techniques in the semiconductor industry. The proposed quantum-well structure is expected to be especially robust since it does not require an external power supply to open and maintain a band gap, and the cladding layers provide protection against environmental degradation of the graphene layer in its device applications.

Graphene's unique electronic band structure has lead to fascinating phenomena such as massless Dirac fermion physics[1-2] and quantum Hall effects[3]. The intriguing physics and potential for device applications have attracted great research interests in recent years. In particular, the extremely high mobility and the easy control of charge carriers have made graphene promising materials for next generation electronics[4]. There is, however, a major challenge to use graphene in electronic applications because of the lack of an energy gap in its electronic spectra[5]. This, for example, prevents the use of graphene in making transistors. With one more graphene layer added, bilayer graphene has different but equally interesting electronic properties. The inversion symmetric AB-stacking bilayer graphene is still a zero-band-gap semiconductor in its pristine form. A vertically applied electric field has been demonstrated to be effective in opening an electronic band gap in bilayer graphene[6], but a downside of this strategy is that it requires an external power supply to maintain the gate bias. Another established approach to opening a gap in graphene is the epitaxial growth of graphene on a substrate or chemical functionalization that changes the Coulomb potential on different sides of the graphene layer[7–10]. All these strategies achieve the band gap opening by breaking the inversion symmetry of the graphene lattice.

Recent study of topological insulators (TIs) indicates that strong spin-orbit coupling (SOC) also can open a band gap in graphene[11]. Intrinsic SOC in pristine graphene, however, induces a band gap that is too small ($10^{-3}$ meV)[12] to be observable. Subsequent work suggested enhancing the SOC in the graphene lattice by heavy adatom doping[13-14], where a gap of several meV's could be achieved; but it is accompanied by other undesirable effects of doping due to considerable electron transfer. An alternative approach invokes the proximity effect from strong TIs that can induce a large SOC in graphene[15-16]. Recently, high-quality heterostructures between graphene and TIs have been synthesized, where a van der Waals epitaxy of ultrathin films of TI materials $Bi_2Se_3$[17-18] or $Sb_2Te_3$[19] were grown on a pristine monolayer or bilayer graphene substrate using the vapor-phase deposition method. We have shown that a monolayer graphene sandwiched between two strong SOC materials, such as $Bi_2Te_3$ or $Sb_2Te_3$[20], is a non-trivial TI with a large bulk gap[21-22]. Compared with freestanding monolayer graphene, bilayer graphene possesses more versatile electronic properties. It is highly interesting to explore the proximity SOC effect on the electronic structure of bilayer graphene.

In this work we introduce a new approach to a robust band gap opening in graphene. The key strategy here is to take advantage of the proximity enhanced SOC in graphene by constructing a quantum-well (QW) structure where a bilayer graphene sheet is sandwiched between two single quintuple-layers (QLs) of $Bi_2Te_3$. This structural design preserves the graphene lattice symmetry, which no longer needs to be broken because the new driving mechanism for opening a band gap is based on proximity induced SOC. This structural design maximizes the proximity SOC effect from the cladding layers from both sizes of the bilayer graphene sheet. Another distinct advantage of the QW design compared to existing doping or functionalization schemes is its superior protection against environmental degradation of the graphene layers. First-principles calculations reveal a large band gap of 44 meV in bilayer graphene in the QW structure, and this gap can be further enhanced by compressing interlayer distance. Tight-binding simulations confirm that the gap opening and subsequent increase under compression are driven by the proximity enhanced Kane-Mele (i.e., SOC) interaction[11]. The band gap is also sensitive to the interlayer stacking pattern between the graphene and cladding layers, which provides another controllable way to tune the gap size.

First-principles calculations based on the density functional theory (DFT) were carried out using the Vienna Ab Initio Simulation Package (VASP)[23]. The exchange correlation interaction was treated within the generalized gradient approximation of the Perdew-Burke-Ernzerhof type[24]. Projector augmented wave potentials were employed to represent the ions. The SOC was included in the calculations. To describe the van der Waals (vdW) type interaction between graphene and the TI surface, we employed a semiempirical correction by Grimme's method[25].

We show in Fig. 1 the proposed QW structure where a bilayer graphene sheet is sandwiched between two $Bi_2Te_3$ QLs. The two graphene layers are in AB-stacking (Bernal stacking), which has the lower formation energy[7]. For the relative position between graphene and $Bi_2Te_3$, we examined two stacking patterns with the Te atoms at the hollow center of hexagon carbon rings, see Fig. 1(c), or on top of the carbon atoms. The calculations indicate that the hollow stacking pattern is 34.7 meV lower in energy than the top stacking pattern. We also have shifted the $Bi_2Te_3$ layer relative to the graphene layer and tracked the variation of the total energy and found that the hollow configuration is the most stable one. We thus focus our study on the stable hollow stacking pattern. Since the QW structure preserves the inversion symmetry of the graphene lattice, it facilitates the examination of the parity and topological nature of the electronic properties. The lattice mismatch is relatively small at 2.8% between graphene ($\sqrt{3}\times\sqrt{3}$ surpercell) and $Bi_2Te_3$ (primitive unit cell). We have chosen the experimental in-plane lattice constant of 4.38 Å for $Bi_2Te_3$ and adjusted the lattice constant of graphene accordingly. For brevity, we label the $Bi_2Te_3$/Bilayer-Graphene/$Bi_2Te_3$ and $Bi_2Te_3$/Single-Graphene/$Bi_2Te_3$ QW as TI/BG/TI and TI/SG/TI, respectively, although single QL $Bi_2Te_3$ is no longer a TI due to the interaction between the states on opposite surfaces[26-27].

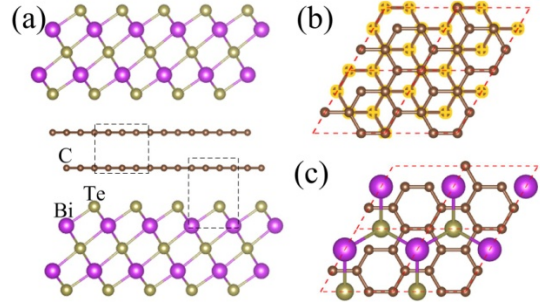

**Figure 1** (a) Side view of the TI/BG/TI quantum-well structure. (b) Top view of the two graphene layers, where the carbon atoms in the bottom layer are highlighted in yellow. (c) Stacking pattern between graphene and $Bi_2Te_3$, where the Te atoms are at the hollow center of the hexagon carbon rings.

The two Dirac valleys of a freestanding bilayer graphene with a $\sqrt{3}\times\sqrt{3}$ supercell located at the K and K′ points are folded to the Γ point in the Brillouin zone, and the Dirac bands are thus four-fold degenerate (excluding spin degeneracy, the valence and conduction bands are each two-fold degenerate). When the bilayer graphene is sandwiched between two $Bi_2Te_3$ QLs, its electronic band dispersion near the Fermi level is not much affected when the SOC is not considered since the Dirac bands are inside the band gap of $Bi_2Te_3$, see Fig. 2(a). Distinct from the freestanding bilayer graphene, which has a zero band gap, a small 2-meV gap is opened. This result is in sharp contrast to the situation of TI/SG/TI QW[22], where the non-SOC band gap of monolayer graphene is opened to 41 meV due to the Kekulé deformation (namely the non-equivalence of the three nearest carbon bond hopping terms)[28]. From Fig. 1(c), one can see that the Kekulé deformation is still present in the TI/BG/TI QW. However, the effect is greatly diminished by the interlayer interaction in bilayer graphene. To demonstrate this point, we artificially increased the distance between two TI-G heterojunctions (i.e., the two halves of the TI/BG/TI QW) and tracked the gap variation, as shown in Fig. 3(a) and 3(b).

The TI/BG/TI with a large distance between the two graphene layers can be regarded as two $Bi_2Te_3$/graphene heterojunctions. The Kekulé deformation produces a gap opening of 5 meV in graphene in the heterojunction. As the distance between the two halves of the structure is gradually reduced, the interlayer interaction in bilayer graphene is enhanced, which produces a considerable band splitting at the $\Gamma$ point. This band splitting largely compensates the gap opened by the Kekulé deformation, leading to the diminished band gap at the equilibrium position of the structure.

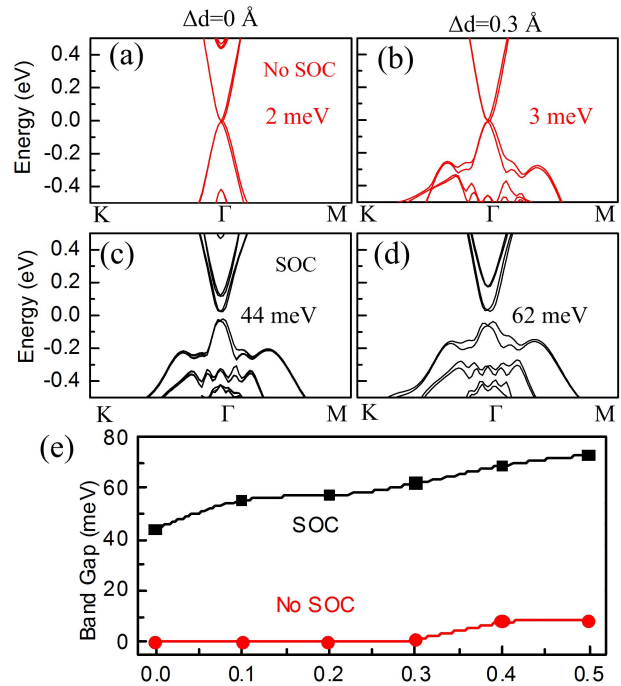

**Figure 2.** Calculated electronic band structure of the TI/BG/TI QW (a) without and (c) with the SOC at the equilibrium position; Presented in (b) and (d) are the corresponding band structures under compression that has reduced the interlayer distance by 0.3 Å. Presented in (e) is the band gap variation as a function of the reduced interlayer distance.

The strong SOC in $Bi_2Te_3$ is expected to produce considerable proximity effect in the graphene layers, thus influence the electronic properties. When the SOC is switched on, the graphene Dirac bands remain inside the $Bi_2Te_3$ gap. A profound SOC effect on the band structure is a significant shift of the graphene Dirac bands away from the Fermi level (see Fig. 2c). It produces a band gap of 44 meV, which represents a four-orders-of-magnitude enhancement over the intrinsic gap of graphene[12] and is large enough for transistor application. Although the $Bi_2Te_3$ QLs play a key role in generating the proximity enhanced band gap in the graphene layer, their electronic bands do not mix near the Fermi level. This is a significant result since it means that the intrinsic high mobility of graphene is essentially unaffected, making the QW structure a desirable system for applications in nanoelectronics. Distinct from the TI/SG/TI case, the proximity SOC enhancement in bilayer graphene does not result in a topological phase transition. As discussed above, the Dirac cones at the $\Gamma$ point of the TI/BG/TI QW are two-fold (excluding the spin degeneracy), which leads to a double band inversion driven by the proximity SOC effect, resulting in a topologically trivial semiconducting phase. This is confirmed by the parity check of the electronic states measured by the $Z_2$ index [29]. We have calculated the $Z_2$ index for the TI/BG/TI QW structure by directly evaluating the parity eigenvalues following the Fu- Kane criteria [29], and found that the product of all the valence band parities at the two time-reversal invariant k-points, $\Gamma$ and *M*, are "+", indicating $Z_2$=0, thus confirming that the QW structure is a topologically trivial semiconductor, where the edge states of the two graphene layers annihilate each other by the interlayer coupling[30], a situation similar to that in some honeycomb compounds [31-32]. Another distinctive feature of the TI/BG/TI QW is the presence of the Rashba effect resulting from the stacking sequence, which can be seen in the band structure where the folded Dirac cones have an obvious shift relative to each other (the effect is more obvious under compression as shown in Fig. 2d). It is worth noting that the proximity SOC (Kane-Mele term) is the dominate effect in determining the electronic structure here, while the Rashba term only leads to spin flip with no influence on the gap value.

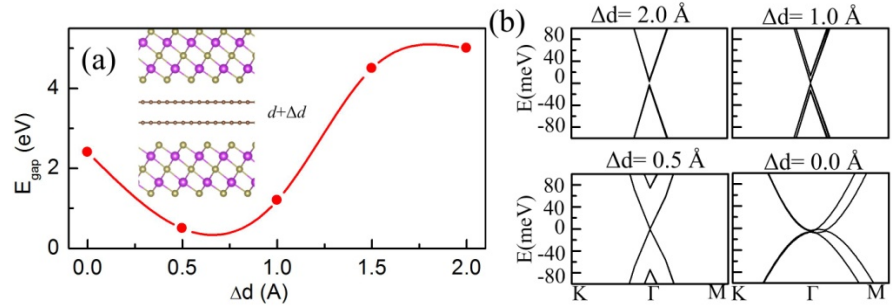

**Figure 3.** (a) The band gap variation with increasing distance between the two graphene layers without considering the SOC; (b) the band structure evolution at different interlayer distances.

The proximity effect is expected to be sensitive to the distance between the graphene and cladding layers. It should be enhanced by reducing the distance between the bilayer graphene sheet and the cladding $Bi_2Te_3$ QLs, which, in turn, could increase the band gap. To prove this idea and provide a feasible approach to effective band engineering, we have simulated interlayer compression by reducing the distance between the upper and bottom Te atoms and then relaxing the structure while fixing the outermost surface atoms. The calculations indicate that the non-SOC band structure is almost unchanged compared to that in equilibrium position except a lifting of the band states of $Bi_2Te_3$. The gap increases to about 3 meV when the distance is reduced by 0.3 Å, see Figure 2(b). However, the SOC gap is significantly increased to 62 meV, which is accompanied by a more obvious Rashba effect that can be seen from the shift of the Dirac cones near the

Fermi level, as shown in Fig. 2d. The SOC band gap undergoes an almost linear increase under compression, while the non-SOC gap is largely insensitive to interlayer distance, see Fig. 2e. In stark contrast, the freestanding bilayer graphene is always metallic regardless of interlayer distance compression. These results not only demonstrate that compression is a feasible way to tune the band gap of bilayer graphene in the QW structure, but also indicate that the gap opening originates from the proximity SOC effect.

To unveil the mechanism and the underlying physics, we performed a TB study on the bilayer graphene and adjusted the parameters to simulate the effects of the bonding environment in the QW structure and the external compression. We write the Kane-Mele Hamiltonian[11,33] for the bilayer graphene lattice including Rashba effect as,

$$H = H_{BG} + H_{K-M} + H_{Rashba} \quad (1)$$

$$H_{BG} = -t \sum_{<ij>,m,\sigma} a^\dagger_{im\sigma} b_{jm\sigma} - \gamma_1 \sum_{<ij>,\sigma} a^\dagger_{i1\sigma} b_{j1\sigma} - \gamma_3 \sum_{<ij>,\sigma} a^\dagger_{i2\sigma} b_{j2\sigma} + H.c. \quad (2)$$

$$H_{K-M} = i\lambda_{SO} \sum_{<ij>,m} v_{ij} (a^\dagger_{im} s^z a_{jm} + b^\dagger_{im} s^z b_{jm}) \quad (3)$$

$$H_{Rashba} = i\lambda_R \sum_{<ij>,m} a^\dagger_{im} (\vec{s} \times \vec{d}_{ij}) a_{jm} + b^\dagger_{im} (\vec{s} \times \vec{d}_{ij}) b_{jm_z} \quad (4)$$

The first term in Eq. (2) describes the hopping in each graphene layer, and in a nearest-neighbor model for freestanding graphene $t$=2.60 eV [34]. The second and third term describe the coupling between the two graphene layers, where $\gamma_1$ =0.4 eV ($\gamma_3$ =0.3 eV) [35] is the coupling strength between the A-sublattice (B-) of the upper layer with the A-sublattice (B-) of the bottom layer. Due to the diminishing non-SOC gap by the interlayer coupling, the Kekulé deformation term is neglected here. In pristine graphene, Kane-Mele interaction $\lambda_{SO}$ is about $10^{-3}$ meV, but here it is significantly enhanced by the proximity effect; the parameters $v_{ij}$=-$v_{ji}$=±1 depend on the direction of electron hopping from site $i$ to $j$: $v_{ij}$=+1 if the electron makes a left turn to a nearest-neighbor bond, and it is

−1 if it makes a right turn. In $H_{Rashba}$, $\vec{s}$ is the Pauli matrix and $s^z$ is the z-compound; $\vec{d}_{ij}$ is the vector linking two next-nearest neighbors.

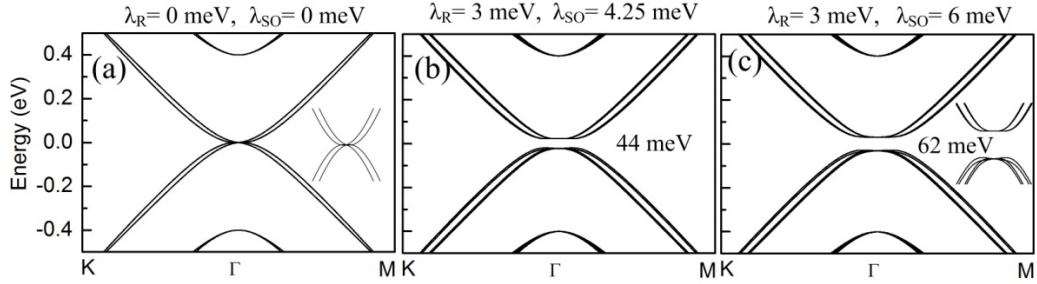

**Figure 4.** Tight-binding modeling results. (a) Metallic state of the bilayer graphene without SOC; (b) gap opening when turning on the Kane-Mele and Rashba term; (c) enhanced gap opening with increasing $\lambda_{SO}$, which corresponds to compressed interlayer distance.

We first examine the electronic band structure without SOC, as shown in Fig. 4a. The same as previously reported[36], the AB-stacked pristine bilayer graphene is a metal with four-fold band degeneracy at the $\Gamma$ point (excluding the spin degeneracy). Once the SOC is turned on, the degenerate Dirac cones at the $\Gamma$ point is lifted by an effective SOC induced repulsive interaction, opening a gap in the bilayer graphene. Due to the stacking sequence as discussed above, the Rashba effect is present, which flips the band states in the k-space. We fixed $\lambda_R$ to 3 meV in our TB simulations because of its weak effect on the gap value. When $\lambda_{SO}$ reaches 4.25 meV, the gap is opened to 44 meV as shown in Fig. 4b, which corresponds to the results of the QW structure at the equilibrium position, see Fig. 2d. Further increasing SOC (e.g., via reducing the interlayer distance) opens a larger band gap, and at the SOC value of 6 meV, the band gap reaches 62 meV (Fig. 4c), corresponding to the result by the ab initio calculations in the compressed structure, see

Fig. 2e. At the graphene Dirac point, the Kane-Mele term (SOC) produces a band gap $(6\sqrt{3}\lambda_{SO})$[14], which explains that compression enhanced SOC would lead to an increased gap.

Although the symmetric structure with the Te atoms at the carbon hexagonal center is energetically favorable, it is instructive to examine structures that have a mismatched interlayer stacking pattern, which could happen during the synthesis either by chance or by design. To check the effect of stacking pattern on the band gap opening, we shift the bilayer graphene relative to the cladding layer and track the corresponding gap variation. The results in Fig. 5a show that both the total energy and interlayer distance between graphene and $Bi_2Te_3$ increase with the interlayer shift. This result indicates that the symmetric QW configuration is stable against interlayer shift and that, because of the increased interlayer distance, the proximity SOC in graphene would become weaker and the band gap would be reduced in QW structures with mismatched stacking patterns. Our calculated results presented Fig. 5b indeed show that the SOC gap decreases with the distance of the interlayer shift. It should be noted that the proximity SOC reduction is not the only mechanism affecting the band gap in the present system. Results in Fig. 5b shows that the non-SOC gap actually increases with the interlayer shift and can reach up to 10 meV, which is attributed to intervalley scattering as a result of crystal field stabilization energy[36]. The competition between the intervalley scattering and proximity SOC mechanisms produces the final gap value, which is determined as the difference between the non-SOC gap and the proximity SOC opened gap[21]. These results indicate that the gap opening in the QW structure is sensitive to the interlayer stacking pattern, which helps understand the operating mechanism in the present system, and it also offers a possible approach to tuning the gap value.

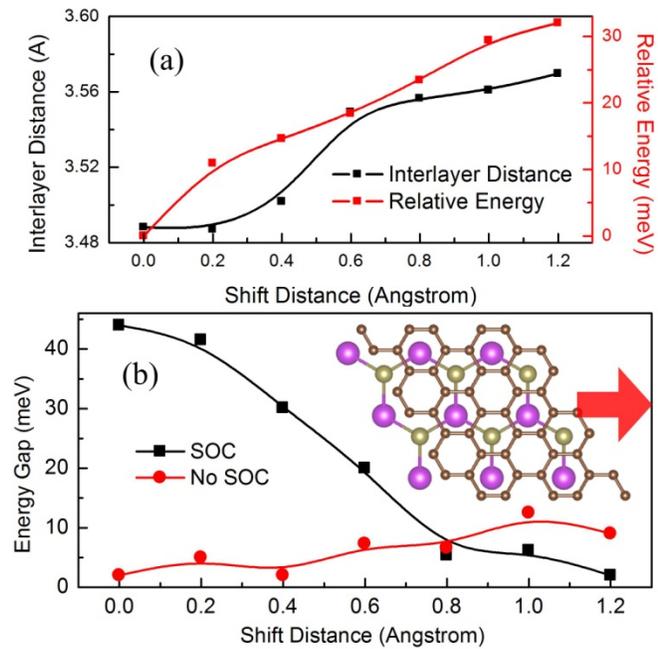

**Figure 5.** (a) Interlayer distance between the graphene and $Bi_2Te_3$ layers and the relative energy as a function of the interlayer shift distance. (b) Energy gap without and with the SOC at different shift distance.

In summary, we have identified by ab initio calculations a new method that opens the band gap of bilayer graphene that is sandwiched between thin slabs of $Bi_2Te_3$. The gap value reaches 44 meV, which is suitable for device applications, and the gap can be further tuned and enhanced by compressing the interlayer distance between graphene and $Bi_2Te_3$. Changing stacking pattern at the interface adds another approach to tuning the band gap. These QW structures can be synthesized by existing epitaxy techniques[17–19], and they are expected to be especially robust because of their structural characteristics. The proposed QW construction offers a promising new platform for the fabrication and design of robust graphene-based nanoelectronics.

## Acknowledgments

Computation was carried out at HLRN Berlin/Hannover (Germany). L.K. acknowledges financial support by the Alexander von Humboldt Foundation of Germany. C.C. was supported by the U.S. Department of Energy under Cooperative Agreement DENA0001982. B.Y. acknowledge financial support from the European Research Council Advanced Grant (ERC 291472). F.H. is supported by the postdoc Research Fellowship from Bremen University.